\documentclass[apjl]{emulateapj}
\usepackage{natbib}
\usepackage{amsmath}
\usepackage{mathptmx}
\usepackage{color}

\usepackage{booktabs}
\usepackage{graphicx}

\newcommand\ho{\ifmmode {\rm HI} \else H{\small I} \fi}
\newcommand\hh{\ifmmode {\rm H_2} \else H$_2$ \fi}
\def\no{\ifmmode {N_{\rm HI}} \else $N_{\rm HI}$ \fi}
\def\nt{\ifmmode {N_{\rm H_2}} \else $N_{\rm HI}$ \fi}
\def\so{\ifmmode {\Sigma_{\rm HI}} \else $\Sigma_{\rm HI}$ \fi}
\def\st{\ifmmode {\Sigma_{\rm H_2}} \else $\Sigma_{\rm H_2}$ \fi}
\def\stot{\ifmmode {\Sigma_{\rm tot}} \else $\Sigma_{\rm tot}$ \fi}
\def\msun{\ifmmode {\rm M_{\odot}}\else $\rm M_{\odot}$\fi}
\def\mpc{\ifmmode {\rm M_{\odot} \ pc^{-2}} \else $\rm M_{\odot} \ pc^{-2}$ \fi}
\def\tra{\ifmmode {\rm HI-to-H_2}\else H{\small I}-to-H$_2$ \fi}

\begin{document}

\title{H{\small I}-to-H$_2$ Transitions in the Perseus Molecular Cloud}

\author{Shmuel Bialy$^{\star}$ $^1$, Amiel Sternberg$^1$, Min-Young Lee$^2$, Franck Le Petit$^3$ and Evelyne Roueff$^3$
}

\altaffiltext{1}
{Raymond and Beverly Sackler School of Physics \& Astronomy, Tel Aviv University, Ramat Aviv 69978, Israel}
\altaffiltext{2}
{Laboratoire AIM, CEA/IRFU/Service d'Astrophyque, Bat 709, 91191 Gif-sur-Yvette, France}
\altaffiltext{3}
{LERMA, Observatoire de Paris, PSL Research University, CNRS, UMR8112, F-92190 Meudon, France}

\slugcomment{Accepted for publication in the Astrophysical Journal}

\begin{abstract}
We use the \citet{Sternberg2014} theory for interstellar atomic to molecular hydrogen
(H{\small I}-to-H$_2$) conversion to analyze \tra transitions in five (low-mass) star-forming 
and dark regions in the Perseus molecular cloud, B1, B1E, B5, IC348, and NGC1333.  
The observed H{\small I} mass surface densities of 6.3 to 9.2 \mpc 
are consistent with \tra transitions dominated by H{\small I}-dust shielding
in predominantly atomic envelopes.
For each source, we constrain the dimensionless parameter $\alpha G$, and 
the ratio $I_{\rm UV}/n$, of the FUV intensity to hydrogen gas density. 
We find $\alpha G$ values from 5.0 to 26.1, implying 
characteristic atomic hydrogen densities
11.8 to 1.8 cm$^{-3}$, for $I_{\rm UV} \approx 1$ appropriate for Perseus.
Our analysis implies that the dusty \ho shielding layers are probably
multiphased, with thermally unstable UNM gas in addition to cold CNM
within the 21 cm kinematic radius.
 
\end{abstract}

\section{Introduction}
Conversion of hydrogen gas from atomic (H{\small I}) to molecular (H$_2$)
form is of critical importance for the evolution of the interstellar medium (ISM) and
for star-formation in galaxies \citep[][hereafter S14]{Sternberg2014}.  

Recently, \citet{Lee2012} and \citet[][hereafter L12/L15]{Lee2015} 
analyzed \tra transitions in several subregions 
within the well-studied Perseus molecular cloud.
The Perseus cloud is itself part of the nearby
Taurus-Auriga-Perseus complex located at a distance of $\sim$300~pc
and is embedded within the Per OB2 \ho supershell \citep{Bally2008}.
The Perseus cloud mass is a few
$10^4 \, \msun$, 
and this includes the \ho plus H$_2$ \citep{Bachiller1986, Kirk2006}.
The cloud consists of an
extended dusty \ho envelope surrounding several condensations of dense H$_2$ gas
\citep{Sancisi1974, Sargent1979, Ungerechts1987, Ridge2006}.
The overall angular size depends on the tracers used, and
a characteristic diameter based on 21 cm kinematics is $\sim 80$~pc
\citep{Imara2011}.

L12/L15 used the \ho data provided by the Galactic Arecibo L-band Feed Array \ho Survey \citep[GALFA-\ho survey; ][]{Peek2011}, together with far-infrared data from the Improved Reprocessing of the IRAS Survey \citep[IRIS; ][]{MivilleDeschenes2005} and the V-band extinction image provided by the COMPLETE Survey \citep{Ridge2006}, to derive \ho and \hh surface densities (\so and $\Sigma_{\rm H_2}$) on $\sim 0.4$~pc scales, for several hundred sight-lines towards five dark and (low-mass) star-forming regions within Perseus.
These are B1, B1E, B5 (dark), IC348, and NGC1333 (star-forming). Here we consider the data 
presented by L15 for which the inferred \ho column densities are corrected for
(small, up to 20\%) 21~cm optical depth effects.
 
L12/L15 analyzed the Perseus data using the formalism presented by
\citet[][hereafter KMT]{Krumholz2009} for interstellar \tra transitions
in optically thick media.
In this paper we use the simpler and more general theory presented by S14 
to reanalyze the Perseus observations.  
In \S~1, we briefly summarize the relevant S14 formalism.
In \S~2 we present and fit the L15 observations of the
$\Sigma_{\rm{H_2}}/\Sigma_{\rm HI}$ ratios 
in Perseus. In \S~3 we use the observed
maximal \ho mass surface densities towards each H$_2$ cloud to 
constrain the controlling dimensionless parameter $\alpha G$ and the
characteristic H{\small I} gas densities in the atomic envelopes.
The observed H{\small I} mass surface densities 
are consistent with \tra transitions dominated by H{\small I}-dust shielding.
The relatively low gas densities we infer suggests that the \ho shielding layers
are probably multiphased and are not pure CNM.

\section{Theory}
\label{sec: theory}

\subsection{HI Column Density}

S14 presented a general analytic formula for the steady state column density of photodissociated \ho gas in optically thick clouds illuminated by FUV radiation,
derived for planar geometry and uniform density gas.
For irradiation by isotropic fields, the {\it total}
\ho column density is given by
\begin{figure*}[!hbt]
\includegraphics[width=.9\textwidth]{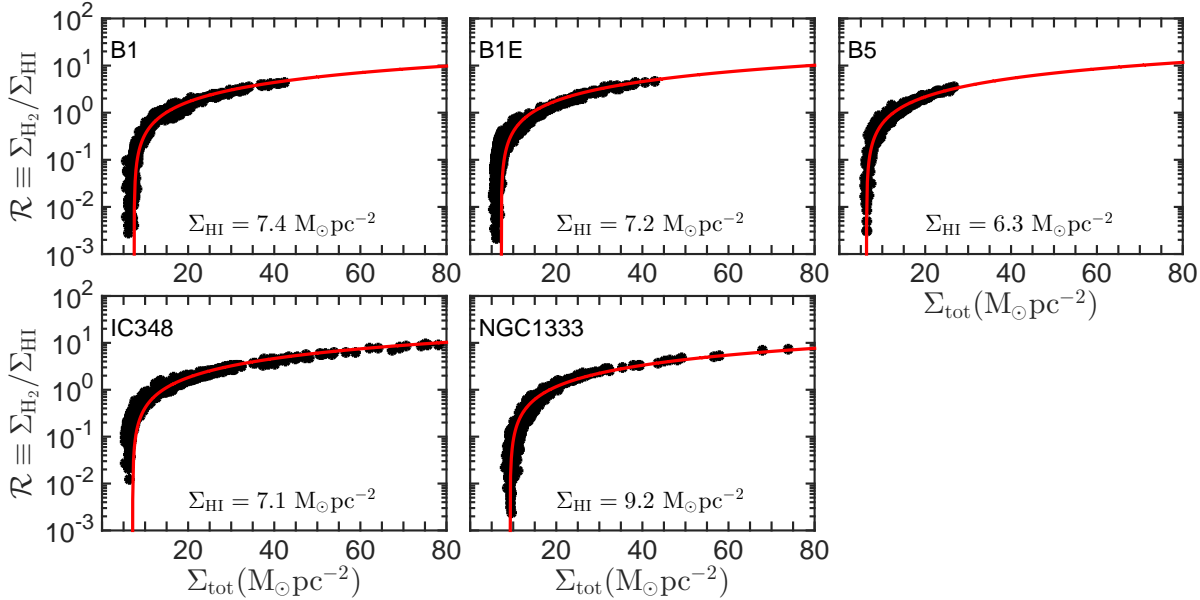}
\caption{$\mathcal{R}_{\hh} \equiv \st/\so$ as a function of $\stot \equiv \so + \st$ for the five regions: B1, B1E, B5, IC348 and NGC1333. The red curves are the ``best-fits" to the data. The median errors (not displayed) are $\pm 15 \%$ and $\pm 35 \%$ in \stot and $\mathcal{R}_{\hh}$ respectively.
The best-fitting values of \so are also indicated.}
 \label{fig: fits}
 \end{figure*} 

 \begin{equation}
\label{eq: HI}
\no \, = \, 2  \times  \frac{\left< \mu \right>}{\sigma_{g}} \, \ln\Big[\frac{1}{\left< \mu \right>} \frac{\alpha G}{4} + 1 \Big] \ .
\end{equation}
In this expression, the factor of two is for two-sided illumination, $\left< \mu \right>=0.8$ is a geometrical factor for isotropic radiation, $\sigma_g$ is the 
dust-grain absorption cross section per hydrogen nucleon
for 912-1108~\AA~Lyman-Werner (LW) band radiation,
and $\alpha G$ is the basic dimensionless parameter.
As in S14 we assume that 
\begin{equation}
\label{eq: sigma_g}
\sigma_{g} \, = \, 1.9 \times 10^{-21} \, \phi_{g} \,  Z'_{g} \ {\rm cm}^2 \ ,
\end{equation} 
where $Z'_{g}$ is the grain abundance relative to standard Galactic ISM grain abundances.  Thus, $Z'_{g}$ is the ``dust-metallicity".  
In Equation~(\ref{eq: sigma_g})
$\phi_{g}$ is a factor of order unity depending on the 
specific grain composition, and the scattering and absorption properties.
The dimensionless parameter 
\begin{equation}
\label{eq: alpha_G_basic}
\alpha G \ \equiv \ \frac{D_0G}{Rn} \ = \ {{\bar f}}_{\rm diss}\frac{\sigma_{g} w F_0}{Rn} \ ,
\end{equation}
where $D_0$ is the free-space H$_2$ dissociation rate (s$^{-1}$), 
$G$ is the average H$_2$ self-shielding factor, $R$ is the H$_2$ formation 
rate coefficient (cm$^3$~s$^{-1}$),
$n$  is the total hydrogen gas density (cm$^{-3}$),
${\bar f}_{\rm diss}=0.12$ is the mean dissociation probability per
H$_2$-absorbed LW band photon, 
$F_0$ is the free-space LW photon flux (cm$^{-2}$~s$^{-1}$), 
and $w\equiv 1/[1+(2.64 \phi_g Z_g')^{1/2}]$ is the normalized H$_2$-dust-limited 
dissociation bandwidth (see S14 for a detailed discussion of all these quantities).  
Physically, $\alpha G$ is the ratio of the H{\small I}-dust absorption rate
of the effective unattenuated H$_2$ dissociation flux, to the H$_2$ formation rate.

 \begin{table}
\caption{Total HI surface densities.}
\centering 
\begin{tabular}{l c c}
\hline\hline 
Cloud &  $\; \Sigma_{\rm HI}$ (\mpc) & $\;  N_{\rm HI}$ (10$^{20}$ cm$^{-2}$)  \\ [0.5ex] 
  \hline 
B1  & 7.4 & 9.3 \\
B1E & 7.2 & 9.0 \\
B5 & 6.3 & 7.9 \\
IC348 & 7.1 & 8.9 \\
NGC1333 & 9.2 & 11.6 \\

\hline 
\end{tabular}
\label{table: sig_HI} 
\end{table}

For H$_2$ formation on dust grains $R = 3 \times 10^{-17} Z'_g$~cm$^{3}$~s$^{-1}$, and the dimensionless parameter $\alpha G$ can be expressed as
\begin{equation}
\label{eq: alpha_G}
\alpha G \ \equiv  \ 1.54 \ \frac{I_{\rm UV}}{n/100 \ {\rm cm^{-3}}} \ \frac{\phi_{g}}{1+(2.64 \phi_{g} Z'_{g})^{1/2}} \ ,
\end{equation} 
where $I_{\rm UV}$ is the field intensity relative to the mean \cite{Draine1978}
interstellar field, such that $F_0=2.07\times10^7 I_{\rm UV}$~cm$^{-2}$~s$^{-1}$ and $D_0=5.8\times 10^{-11} I_{\rm UV}$~s$^{-1}$. 
The value of $\alpha G$ determines the nature of the \tra transition
and the size of the integrated \ho column.
In the ``weak-field" limit $\alpha G \ll 1$, $N_{\rm HI}$ is small, and the \tra 
transition is controlled by H$_2$-line and H$_2$-dust absorption.
In the ``strong-field" limit $\alpha G \gg 1$, $N_{\rm HI}$ is large, and the \tra transition is dominated by H{\small I}-dust absorption. Importantly,
for optically thick clouds the total H{\small I} column density, $N_{\rm HI}$, depends only on $\alpha G$ and $\sigma_{g}$.

\subsection{Time-Scale}
Equation (\ref{eq: HI}) for the \ho column density is for steady-state conditions such that the local H$_2$ destruction rate equals the formation rate, at every location. The equilibrium time-scale for \ho/\hh formation-destruction is
\begin{equation}
\label{eq: time scale}
t_{\rm eq} \ = \ \frac{1}{D \ + \ 2 \ R \ n} \ ,
\end{equation} where $D$ is the local  (attenuated) photodissociation rate.
For molecular gas $D/(2 R n) \ll 1$ and $t_{\rm eq} \simeq 1/(2 R n) \approx 5\times10^8/n$~yr.
For atomic gas $D/(2Rn) \gg 1$ and $t_{\rm eq} \simeq 1/D$. In free-space $D=D_0$ and $t_{\rm eq} \approx 5.5 \times 10^2/I_{\rm UV}$~yr.

\subsection{Multiphased Gas}

For a multiphased CNM/WNM mixture of H{\small I} gas
in which the heating is dominated by photoelectric emission 
from dust grains, the field intensity $I_{\rm UV}$ and the density $n_{\rm CNM}$ of the CNM are correlated, with
\citep{Wolfire2003}
\begin{equation}
\label{eq: nCNM}
n_{\rm CNM}=22.7 \, I_{\rm UV} \Big(\frac{4.1}{1+3.1Z'^{0.365}_g} \Big) \Big( \frac{Z'_g}{Z'_{\rm C,O}} \Big) \Big( \frac{\phi_{\rm CNM}}{3} \Big) \ {\rm cm^{-3}} \  .
\end{equation}
Here $Z'_{\rm C,O}$ is the {\it gas phase} carbon-oxygen abundance
relative to the abundances at solar metallicity. That is, 
$Z'_{\rm C,O}$ is the ``gas-phase metallicity". 
In Equation~(\ref{eq: nCNM}), $\phi_{\rm CNM}$
expresses
the range for which the CNM can
be in pressure equilibrium with the WNM. Typically, $\phi_{\rm CNM}\sim 3$.
For multiphase conditions the WNM density is $n_{\rm WNM} \sim 0.01 n_{\rm CNM}$,
and the gas is thermally unstable (UNM) for densities between $n_{\rm WNM}$
and $n_{\rm CNM}$ (see Figure 9 of \citealt{Wolfire2003}).
 
It follows from Equations (\ref{eq: alpha_G}) and (\ref{eq: nCNM}) that for 
H{\small I}-to-H$_2$ transitions occurring in pure CNM at thermal pressures allowing multiphase conditions (KMT, S14)
\begin{align}
\label{eq: aGCNM}
\alpha G = (\alpha G)_{\rm CNM} &\equiv 2.58 \Big( \frac{1+3.1 Z_{g}'^{0.365}}{4.1} \Big) \Big( \frac{Z'_{\rm C,O}}{Z'_{g}} \Big)\\\nonumber
&\times \Big(\frac{3}{\phi_{\rm CNM}}\Big) \Big(\frac{2.62}{1+(2.64 \phi_{g} Z'_{g})^{1/2}} \Big) \phi_{g} \; .
\end{align}
In \S~4 we will consider perturbations to $(\alpha G)_{\rm CNM}$ by varying
independently the parameters $\phi_g$, $Z^\prime_g$, and $Z^\prime_{\rm C,O}$, all
for $\phi_{\rm CNM}$ in the realistic range of 2 to 5.

\section{$\Sigma_{\rm H_2}/\Sigma_{\rm HI}$ in Perseus}
\label{sec:Fitting}

Equation (\ref{eq: HI}) can be reexpressed as an \ho mass surface density
\begin{equation}
\label{eq: Sigma_HI}
\so \, = \, 6.71 \ \Big( \frac{1.9}{\sigma_{g -21}} \Big) \, \ln\Big[ \frac{\alpha G}{3.2} + 1 \Big] \  \mpc \ ,
\end{equation}
where $\sigma_{g -21}=\sigma_{g}/(10^{-21}$~cm$^2)$.
(In Equation~(\ref{eq: Sigma_HI}) the contribution of helium to the
mass is not included).
If $\stot \equiv \so + \st$ is the total hydrogen gas surface density, and
$\mathcal{R}_{\rm H_2} \equiv \st/\so$ is the molecular-to-atomic mass ratio,
then
\begin{equation}
\label{eq: R}
\mathcal{R}_{\rm H_2}(\stot) \, = \, \frac{\stot}{\so} \, - \, 1 \ .
\end{equation} 
For optically thick clouds, 
$\so(\sigma_{g},\alpha G)$ (as given by Equation [\ref{eq: Sigma_HI}])
is independent of $\st$, and 
$\mathcal{R}_{\rm H_2}$ varies linearly with \stot with slope $1/\so$.

 \begin{figure*}[!hbt]
\centering
 \includegraphics[width=1.0\textwidth]{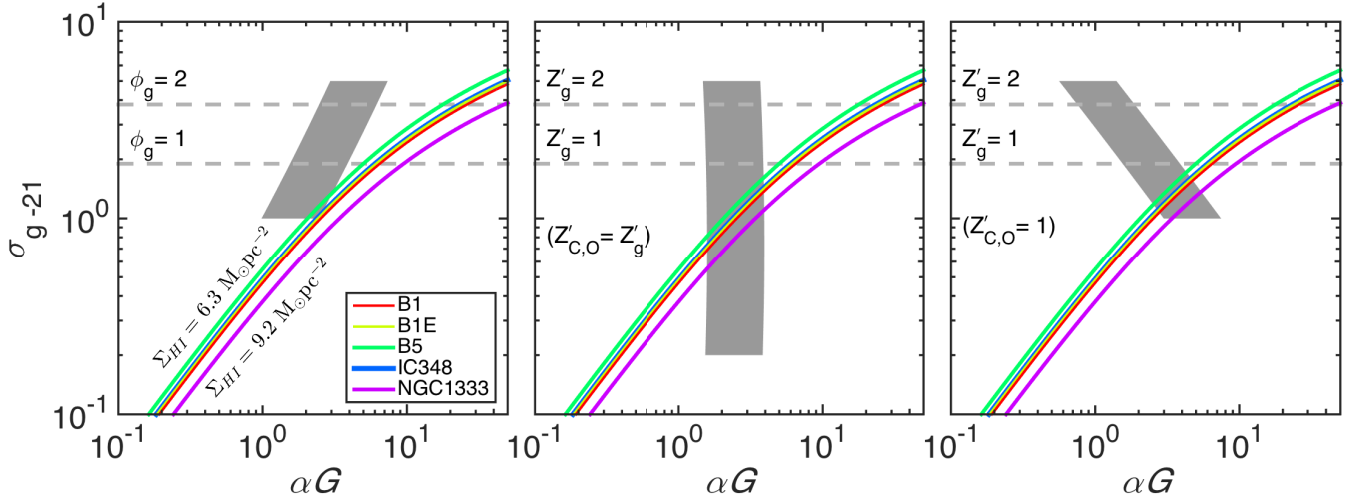}
\caption{The observed HI contours in the $\sigma_{g} - \alpha G$ parameter space.  
The horizontal dashed lines are for $\sigma_{g-21}=1.9$ and 3.8.
The grey strips are where $\alpha G = (\alpha G)_{\rm CNM}$ (Equation [\ref{eq: aGCNM}]) for $\phi_{\rm CNM}$ in the range $2 - 5$, (with $\phi_{\rm CNM}$ increasing from right to left across the strip).
The dust cross section $\sigma_{g-21}=1.9 \phi_g Z'_g$ varies with the dust abundance $Z'_g$ and the intrinsic dust absorption properties $\phi_{g}$. The left panel is for variations in $\phi_{g}$ assuming $Z'_{g}=Z'_{\rm C,O}=1$. 
The middle and right panels are for variations in $Z'_{g}$ with $\phi_{g}=1$, assuming $Z'_{\rm C,O}=Z'_{g}$ (middle) or $Z'_{\rm C,O}=1$ (right), see \S 4 for details.
}
 \label{fig: sig_alphaG}
 \end{figure*}
 
In Figure \ref{fig: fits} we plot the L15 data for $\mathcal{R}_{\rm H_2}$ versus \stot
for the five dark and star-forming regions in Perseus B1, B1E, B5, IC348 and NGC1333. 
A complete discussion of the data extraction methodology is presented
in L12/L15.
For each region, each data point corresponds to a distinct sight-line through
the complex.  Many of the sight-lines probe substantial columns of H$_2$
and pass well through the \tra transition layers.
For example, $\mathcal{R}_{\rm H_2}$ approaches 10 in IC348 and NGC1333.


We fit our Equation (\ref{eq: R}) to the data points
using a standard weighted-least-squares procedure, 
and find the best-fitting total \ho surface density for each region.
The total \ho surface mass densities lie within the narrow range
of 6.3 to 9.2 $\mpc$, and are listed in Table \ref{table: sig_HI}.

The (red) curves in 
Figure \ref{fig: fits} are our best fits for $\mathcal{R}_{\rm H_2}$ versus \stot
as given by Equation (\ref{eq: R}) and the \ho surface densities in Table \ref{table: sig_HI}.
The theoretical curves are in excellent agreement with the data. 
This implies that the
sight-lines are indeed probing optically thick complexes
with complete \tra transitions.

For characteristic CNM, $\alpha G = (\alpha G)_{\rm CNM} \approx 2.58$ as given by Equation (\ref{eq: aGCNM}). For a standard $\sigma_{g-21}=1.9$ the \ho column for pure CNM shielding is then $\Sigma_{\rm HI}^{\rm CNM}=4.0 \ {\rm \msun \ pc^{-2}}$,
significantly smaller than the observed total \ho columns.
This implies that if the entire \ho columns are contributing to the shielding of the 
H$_2$ cores in Perseus, these shielding columns must be multiphased, not just CNM. 
Alternatively, if the shielding is assumed to be entirely CNM, not all of the observed \ho contributes to the shielding.
We discuss these conclusions in more detail in \S4.

\section{Analysis}
\label{sec:Analysis}
\subsection{Shielding and \ho Gas Densities}

According to Equation~(\ref{eq: HI}) or (\ref{eq: Sigma_HI}) the total H{\small I}
surface density depends on just $\sigma_{g}$ and $\alpha G$,
so that curves of constant \ho surface density may be drawn in the
$\sigma_{g}$ versus $\alpha G$ parameter space.
In Figure \ref{fig: sig_alphaG} (all three panels) we plot the locus curves (in color)
for the $\Sigma_{\rm HI}$ inferred for each of the five Perseus regions.  As expected, 
for any $\Sigma_{\rm HI}$ a large $\alpha G$ requires a large $\sigma_{g}$, and vice versa.

The two horizontal dashed lines in Figure \ref{fig: sig_alphaG} represent the 
range of grain absorption cross sections we consider in our analysis, from the standard
$\sigma_{g -21} = 1.9$ to a larger $\sigma_{g -21}=3.8$.
We consider an enhanced $\sigma_g$ 
because L15 found that the visual extinction per hydrogen gas column in Perseus is
$A_V/\no=1.0 \times 10^{-21} \ {\rm mag \ cm^2}$, about a factor of 2 larger than for standard Galactic extinction, and this may imply a correspondingly larger than usual dust-grain absorption cross section.
A larger $\sigma_g$ could be due to (a) altered dust properties at a normal dust-to-gas mass ratio, i.e.~$\phi_g=2$ and  $Z^\prime_g=1$,
or (b) typical (diffuse) ISM dust but with a higher abundance, 
i.e~$\phi_g=1$ and $Z^\prime_g=2$. In any case, 
the gas-phase metallicity, $Z^\prime_{\rm C,O}$,
appears close to Solar in Perseus \citep{Hernandez2009}.


  \begin{table*}[]
\caption{$\alpha G$, volume density ranges, and length scales, for $\sigma_{g-21}$ in the range $1.9-3.8$ for $I_{\rm UV}=1$.}
\centering 
\begin{tabular}{l c c@{\hspace{.8cm}} c c c@{\hspace{.8cm}} c c}
\hline\hline 
&  & & \multicolumn{2}{c}{$\phi_g=1 - 2$ with $Z'_g=1$}  & & \multicolumn{2}{c}{$Z'_g=1 - 2$ with $\phi_g=1$} \\ \cmidrule(lr){4-5} \cmidrule(lr){7-8}
Source &  $\alpha G \; \; \; $ & & $n$ (cm$^{-3}$) & Length Scale (pc) & & $n$ (cm$^{-3}$) & Length Scale (pc) 
  \\ [0.5ex]
   \hline
B1  & 6.5 --  26.1 & & 9.1 -- 3.6  &  33 -- 84 & & 9.1 -- 1.8 &  33 -- 169 \\
B1E  & 6.1 -- 23.8 & & 9.6 -- 3.9 & 30 -- 74 & & 9.6 -- 2.0 & 30 -- 149 \\
B5  & 5.0 --  17.7 & & 11.8 -- 5.3 & 22 -- 49 & & 11.8 -- 2.6 & 22 -- 97 \\
IC345  & 6.0 -- 23.2 & & 9.8 -- 4.0 & 30 -- 72  & & 9.8 -- 2.0 & 30 -- 144 \\
NGC1333  & 9.5 -- 47.0 & & 6.2 -- 2.0 & 61 -- 189 & & 6.2 -- 1.0 & 61 -- 379 \\ 
\hline 
\end{tabular}
\label{table: alpha G densities} 
\end{table*}

For any assumed $\sigma_g$ the implied $\alpha G$ for each source
may be read off the plots in Figure \ref{fig: sig_alphaG}.  In Table \ref{table: alpha G densities}
we list the range of inferred $\alpha G$ parameters for each region, for 
$\sigma_{g -21}=1.9$ to 3.8.
The $\alpha G$ are large ($\gtrsim 1$) and this implies
that the attenuation
of the photodissociating LW radiation is in the strong-field limit
with \tra transitions dominated
by dust absorption within the outer atomic envelopes (``\ho-dust").
This as opposed to \tra transitions controlled by
H$_2$-line self-shielding.

For any given $\alpha G$, the effective gas densities, $n$, in the \ho gas depends on the assumed FUV radiation intensity $I_{\rm UV}$
(see Eq.~\ref{eq: alpha_G}). 
By ``effective" we mean for a uniform density medium.
As discussed by L12 within most of the Perseus system, the 
photodissociating radiation is dominated by the background 
Galactic light.  This is consistent with the overall thermal infrared dust emission
temperatures ($16-22$~K) as well as with the anomalous microwave emissions
\citep{Tibbs2011}. For FUV dust heating, $I_{\rm UV}\approx 1$ 
within a factor of 2.
The radiation fields
near IC345 and NGC1333 may be locally enhanced by the
presence of one or two B5 V-type stars.

In Table \ref{table: alpha G densities}
we list the inferred gas densities
assuming $I_{\rm UV}=1$, for $\sigma_{g -21}=1.9$ and 3.8.
The inferred densities scale linearly with the assumed $I_{\rm UV}$.
For $\sigma_{g -21}=3.8$ the densities depend on whether
(a) $\phi_g=2$ and $Z^\prime_g=1$, or (b)
$\phi_g=1$ and  $Z^\prime_g=2$. The densities are a factor-two
smaller for the second option (see again Eq.~\ref{eq: alpha_G}).
Overall the effective \ho densities range from $\sim 2$ to 10~cm$^{-3}$.
The gas densities in the H$_2$ cores are likely larger, enabling
molecule formation on a time scales 
 $1/(2Rn) \simeq 5\times10^8/n$ yr (see Equation (\ref{eq: time scale})),
within the lifetime of the Perseus cloud $\sim 10 - 100$ Myr.
In the \ho layers the equilibrium time-scales are $1/D \ll 1/(2Rn)$ and a photodissociation steady state is achieved.

In Table \ref{table: alpha G densities}
we also list the range of length scales, ${\ell}\equiv N_{\rm HI}/n$, 
for the atomic shielding envelopes given the observed H{\small I} columns.
With exception of NGC 1333 (which may be influenced
by a B5V star) the derived length scales
are comparable, and within factors 2-3, with the overall 
$\sim 80$~pc \ho kinematic size scale of the Perseus complex.
The inferred sizes are perhaps too large for ``option-b"
($Z'_g=2$, $\phi_g=1$) and more consistent with 
``option-a" ($Z'_g=1$, $\phi_g=2$) for which a larger
dust cross section reflects an intrinsic variation in 
grain properties.

\subsection{Is the H{\small I} Multiphased?}

The gas densities that we have inferred above are lower
than the densities expected for pure CNM as given by
Equation~(\ref{eq: nCNM}), and are intermediate between
$n_{\rm CNM}$ and $n_{\rm WNM}$.  This suggests
that the observed \ho columns are multiphased
mixtures.  If most of the \ho extends to just the
kinematic diameter of $\sim 80$~pc \citep{Imara2011} 
much of the \ho must be thermally unstable, possibly
in a cooling transition from the WNM to CNM phases.

These conclusions are also indicated by the positions of the grey
strips in Figure \ref{fig: sig_alphaG}. The strips show the regions
in the parameter space for which $\alpha G=(\alpha G)_{\rm CNM}$
for different (realistic) perturbations of the controlling quantities 
in Equation~(\ref{eq: aGCNM}).  
In the left-hand panel
the grey strip is for variations in $\phi_g$ from 0.5 to 3 assuming
$Z^\prime_g=Z^\prime_{\rm C,O}=1$. The middle panel
is for metallicities between 0.1 and 3 assuming 
$Z^\prime_g=Z^\prime_{\rm C,O}$ and with $\phi_g=1$.
The grey strip in the right-hand panel is for variations in
just $Z^\prime_g$ from 0.5 to 3, but with 
$Z^\prime_{\rm C,O}=1$ and again $\phi_g=1$.
The width of each strip corresponds to the range
$2 \le \phi_{\rm CNM} \le 5$, for CNM at multiphased conditions,
with $\phi_{\rm CNM}$ increasing from right to left across
the strips.

It is evident from Figure \ref{fig: sig_alphaG} that for
$\sigma_{g -21}$ between 1.9 and 3.8, the contours
for the observed H{\small I} column densities are 
to the right of the grey strips, with $\alpha G$ always
significantly greater than $(\alpha G)_{\rm CNM}$
(i.e.,~$n < n_{\rm CNM}$)
for all types of perturbations in the parameters 
$\phi_g$, $Z^\prime_g$ and $Z^\prime_{\rm C,O}$
shown in the three panels.
This implies that the \ho cannot consist of pure CNM gas.
For the \ho to be fully CNM, the dust absorption cross section
would have to be lower than expected, e.g. if the metallicity were reduced
(see middle panel).

Importantly, our conclusion that the \ho cannot be pure CNM
depends on the inclusion of the denominator
$w\equiv 1/[1+(2.64\phi_g Z_g^\prime)^{1/2}]$ in Equation~(\ref{eq: alpha_G}). As discussed by S14, 
$w$ accounts for the reduction of the effective dissociation bandwidth
by H$_2$-dust.
(This factor is not included in the KMT approximations.)
For example, $w=0.4$ for $\sigma_{g -21}=1.9$
with $\phi_g=1$ and $Z^\prime_g=1$.
Excluding $w$ in Equation~(\ref{eq: alpha_G}) would shift
the ``CNM strips" to the right in Figure \ref{fig: sig_alphaG}
much closer to the \ho contours.  Without the
H$_2$-dust absorption term the inferred effective \ho volume densities would
be more than twice larger than listed in Table \ref{table: alpha G densities},
and much closer to $n_{\rm CNM}$ as given by Equation~(\ref{eq: nCNM})
for $I_{\rm UV}=1$.

\section{Discussion and Summary} 
\label{sec_summary}
L12/L15 used the ``spherical cloud" model developed by KMT
(and updated by \citealt{McKee2010}) to analyze the \tra
transitions in the various Perseus regions (e.g., see Fig.~11 in L15).
In their analysis, each region consists of many individual spheres
with H$_2$ cores surrounded by H{\small I} shells, and
each sight-line probes the area-averaged mass ratios
${\mathcal R}_{{\rm H_2}}\equiv 
\langle\st\rangle/\langle\so\rangle$
as functions of $\langle \stot\rangle$ for each sphere.
Furthermore, L12/L15 adopted the KMT ansatz that the H{\small I}
shielding envelopes are dominated by CNM, and then estimated
$\phi_{\rm CNM}$ for each region assuming $n=n_{\rm CNM}$.

As discussed in detail by S14 the differences in the predicted 
\ho columns and H$_2$ mass fractions for individual clouds
are very small for plane-parallel versus spherical geometries.  
However, the $\chi_{\rm KMT}$ parameter differs from the S14 $\alpha G$
(see also \citealt{Sternberg1988}), since it does not include the
H$_2$-dust absorption factor 
$w\equiv1/[1+(2.64\phi_gZ'_g)^{1/2}]$. 
For a given $\alpha G$ (Equation~[{\ref{eq: alpha_G}])
and an assumed $I_{\rm UV}$ the implied
gas density $n$ is increased if $w$ is excluded. Thus,
including the H$_2$-dust term in $\alpha G$ is essential
for determining how close the gas
density is to the CNM density for multiphased conditions.

In their analysis, L12/L15 assumed that 
$Z'_g=Z'_{\rm C,O}=1$.
They also set
$\sigma_{g -21}=1.0$ corresponding to
$\phi_g=1/1.9$ in our Equation~(\ref{eq: sigma_g}).
It is evident from the left-hand panel of our Figure~\ref{fig: sig_alphaG}
that for $\sigma_{g -21}=1.0$,  the contours for the observed
\ho columns imply $\alpha G $ of 2.0 to 3.4, and are
to the right of the grey CNM strip.  
Enforcing CNM for the \ho would then require $\phi_{\rm CNM} \lesssim 2$.
L12/L15 derived larger $\phi_{\rm CNM}$ ($\sim 5-10$) for the various regions
because the H$_2$-dust term $w$ is not included in the
$\chi_{\rm KMT}$ they used.  Without this factor the grey CNM strips
in Figure~\ref{fig: sig_alphaG} are shifted to the right.
For $\sigma_{g -21}=1.0$ the inferred $\phi_{\rm CNM}$ factors are then increased
to the larger values found by L12/L15.  

In our analysis, we do not assume {\it a priori} that 
$\alpha G=(\alpha G)_{\rm CNM}$, instead we 
infer the effective \ho gas densities. Our inferred densities
range from $\sim 2$ to 10~cm$^{-3}$ depending on
the precise FUV intensity in Perseus, and on the assumed
metallicities and FUV grain absorption cross section.
These densities suggest that the \ho shielding envelopes in Perseus are 
likely multiphased mixtures.  If most of the \ho
is limited to the kinematic 21 cm radius and 
shields the H$_2$ cores, then a significant fraction must be
thermally unstable. Alternatively some of the \ho could be
very extended WNM and not associated with the shielding. 
This follows from just our radiative transfer analysis of the \tra transitions
and \ho columns using our Equation~(\ref{eq: HI}).
The behavior in Perseus suggests that in addition to CNM, 
less dense UNM and perhaps some diffuse WNM, are important in controlling the 
global \tra transitions and Schmidt-Kennicutt thresholds in
external galaxies, from low- to high-redshifts \citep{Leroy2008, Tacconi2010, Bolatto2011, Schruba2011, Genzel2013}.

\begin{acknowledgements}

We thank B.-G. Andersson, A. Goodman and C. F. McKee for helpful conversations about the
Perseus cloud. 
S.B.~acknowledges support from the Raymond and Beverly Sackler Tel Aviv University -- Harvard/ITC Astronomy Program.
M.-Y. L acknowledges supports from the DIM ACAV.
This work was also supported in part by the DFG via German -- Israeli 
Project Cooperation grant STE1869/1-1/GE625/15-1, 
by the PBC Israel Science Foundation I-CORE Program grant 1829/12, 
by the grant SYMPATICO (ANR-11-BS56-0023) from the French Agence Nationale de la Recherche, 
and by the French CNRS national program PCMI.  
\end{acknowledgements}

%
%
%

\end{document}